%% file: godcs.tex
\documentclass[12pt]{iopart}
\usepackage{epsfig}

\usepackage{color}
\def\bea{\begin{eqnarray}}
\def\ena{\end{eqnarray}}

\def\be{\begin{eqnarray}}
\def\en{\end{eqnarray}}

\newcommand\mysql{{\small MySQL\ }}
\newcommand\dbms{{\small DBMS\ }}

\newcommand\godcs{{\small GODCS\ }}
\newcommand\godcss{{\small GODCS}}

\newcommand\geopp{{\small GEO++\ }}
\newcommand\geopps{{\small GEO++}}
\newcommand\cpp{{\small C++\ }}

\newcommand\geo{{\small GEO\,600\ }}

\newcommand\psd{{\small PSD}}

\begin{document}

\title{GEO600 Online Detector Characterization System}
\author{R. Balasubramanian, S. Babak, D. Churches, T. Cokelaer}
\address{School of Physics and Astronomy, Cardiff University, Cardiff CF2 3YB, UK.}

\begin{abstract}
A world-wide network of interferometric gravitational wave detectors is currently operational. The detectors in the network are still in their commissioning phase and are expected to achieve their design sensitivity over the next year or so.
%Since gravitational waves couple weakly to the detector, detecting gravitational wave signals from the noisy output of the interferometer is an extremely challenging data analysis problem.
Each detector is a complex instrument involving many optical, mechanical and electronic subsystems and each subsystem is a source of noise at the output of the detector. Therefore, in addition to recording the main gravitational wave data channel at the output of the interferometer, the state of each detector subsystem is monitored and recorded.
%To take a few examples the fluctuations of laser power in the cavity, the various length sensing feedback and error signals, the auto alignment error and feedback signals, the ambient magnetic fields, seismic motions, wind motions are all measured and recorded.
The analysis of this subsidiary data serves a dual purpose: First,  it helps us to  identify the primary sources of noise which could then be either removed altogether or reduced substantially and second, it helps us in vetoing spurious signals at the output of the interferometer. However, since this subsidiary data is both large in volume (1MB/sec) as well as complex in nature, it is not possible to look at all this data manually. We require an online monitoring and analysis tool which can process all the data channels for various noise artefacts such as transients, drifting of narrowband noise sources, noise couplings between data channels, {\em etc.}, and summarize the results of the analysis in a manner that can be accessed and interpreted conveniently.
%This analysis is computationally demanding in terms of raw computing power as well as memory and storage requirements.

In this paper we describe the \geo Online Detector Characterization System (\godcss), which is the tool that is being used
to monitor the output of the \geo gravitational wave detector situated near Hannover in Germany. We describe the various algorithms that we use and how the results of several algorithms can be combined to make meaningful statements about the state of the detector.
% and how the results of the analyses feed into gravitational wave data analysis.
 We also give implementation details such as the software architecture and the storage and retrieval of the output of \godcss. This paper will be useful to researchers in the area of gravitational wave astronomy as a record of the various analyses and checks carried out to ensure the quality and reliability of the data before searching the data for the presence of gravitational waves.
\end{abstract}
\maketitle

\section{Introduction}
\label{Intro}
\input{Introduction}

\section{The \geopp software library}
\label{GEOSoft}
In this Section we describe the \geopp software out of which \godcs is constructed. The \geopp software has been developed using the \cpp programming language \cite{Stroustrup} and makes extensive use of its object-oriented features. The software is designed to be highly extensible since the task of detector characterization is complex and continuously evolving. The \geopp code relies heavily on the Standard Template Library (STL) \cite{stl} which allows us to define data containers where the type of the data is not known {\em a priori}. The \geopp software is built around a basic templated data container class called {\tt Matrix}.

In Section \ref{dsp} we describe the various software classes that implement core signal processing utilities such as IIR and FIR filters, Fourier transforms, spectrograms {\em etc}. In Section \ref{mysqlframe} we discuss the input/output facilities in \geopps. In Section \ref{geoppmon} we elaborate on the concept of a monitor which is the basic unit of analysis in \geopp. In Section \ref{geopparch} we describe how the various monitors can be run in parallel over a  network of
computers in a client server mode and, finally, in  Section \ref{dbasestruct} we discuss the structure of the database to which the output of \godcs is recorded.
\input{dsp}
\input{geopp}
\section{Monitor Descriptions}
\label{Monitors}
In this Section we describe the various monitors in \godcss. The set of monitors is at present fairly comprehensive in the sense that a wide variety of algorithms have been implemented. The purpose here is to give a summary description of the important monitors and outline their use. Details of the algorithms will be published elsewhere.
\input{PsdMon}
\input{LineMonitor}

\input{StasMonitors}
\input{balaMonitors}
\section{Data Mining using Triana}
\label{Triana}
\input{Triana}

\section{Conclusions}
\label{Conclusion}
\input{Conclusions}

\section*{Acknowledgements}
 This research was supported by the Particle Physics and Astronomy Research Council, UK (grant number PPA/G/O/2001/00485). The authors would like to thank Hartmut Grote, Martin Hewitson, Joshua Smith, Uta Weiland and other members of the \geo experimental group for helping us to learn about the experimental setup of the \geo detector.
The authors would also like to thank Siong Heng and B. S. Sathyaprakash who helped us to validate and 
debug the various \geopp monitors.

\input{References}
\end{document}

%% file: Introduction.tex
%\section{Introduction}

The international network of the ground-based gravitational wave detectors  
consists of three interferometers of the {\small LIGO} project \cite{LIGO},
the {\small VIRGO} detector \cite{VIRGO}, the {\small TAMA\,300} detector \cite{TAMA} 
and the \geo detector
\cite{GEO}. These detectors are searching for gravitational waves 
from a number of different astrophysical sources such as inspiralling close compact
binaries, supernovae explosions, asymmetric rapidly rotating pulsars and relic gravitational waves which are remnants 
of the Big Bang. 
In this paper we will focus our attention on the German-British \geo detector.
% and its online detector characterization system.
\subsection{\geo and subsidary data.}
\begin{figure}[tbh]
\centerline{\epsfysize=8cm \epsfbox{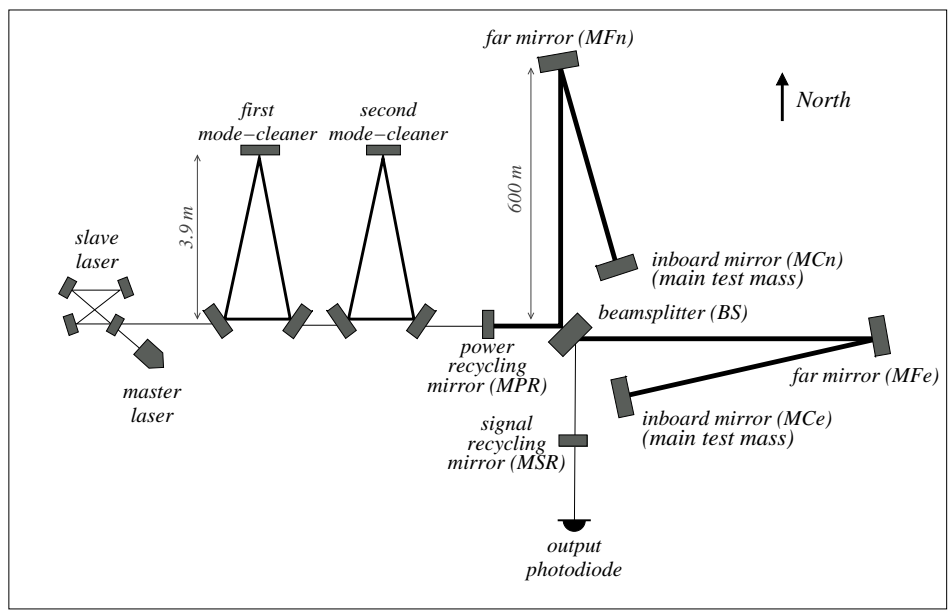}}
\vspace{0.2cm}
\caption{A schematic of the optical layout of \geo showing the input laser system, the two sequential mode cleaners
followed by the dual-recycling Michelson interferometer.}
\label{opticalPath}
\end{figure}

The optical layout of \geo can be divided in three parts: the laser system, the mode cleaners and the dual recycled Michelson interferometer. 
The 12 W laser is spatially filtered by two sequential mode cleaners before being injected into the interferometer. 
The dual recycled \cite{dualRec} Michelson interferometer is used to overcome a 
constraint in the arm length (600m) and to make \geo comparable in sensitivity to the initial kilometer-scale 
interferometers. 
%Power recycling allows us to maximize the light power circulating in the interferometers arms
%(and, therefore reduce photon shot noise) without decreasing the bandwidth of the detector or installing a more 
%powerful laser. The signal recycling in combination with the power recycling is used to optimize the signal 
%storage time, i.e. the average time for which the phase modulated sidebands induced by gravitational wave are stored 
%in the interferometer. This further increases sensitivity of detector. Let us follow the path of light inside the dual-
%recycled interferometer. 
The optical path of laser is shown in Figure \ref{opticalPath}.

The three control loops defined by the Michelson interferometer, 
the power recycling and the signal recycling cavities are used to bring 
and keep the dual-recycled interferometer in lock (in the operational state) 
with the help of feedback loops. 
All optical components (except for the laser and output photodiode) are suspended inside a vacuum 
system. Advanced techniques for seismic isolation \cite{Plissi, Mart} are used in \geo in order to 
reduce the displacement noise of the mirrors introduced by ground motions. 

%Feedback and error point signals are the signals which govern those control loops.
%
%Those loops 
%corresponds to three coupled optical subsystems which are: (i) the Michelson interferometer with its differential and common 
%mode degrees of freedom, (ii) the power-recycling cavity (PRC) formed by the power-recycling mirror and Michelson
%interferometer at its dark fringe operational point, and (iii) the signal-recycling cavity set by the signal-recycling
%mirror and the Michelson interferometer. 

%The experimental components of \geo are installed in three different buildings: the central station and twoend-stations
% to the north and east of the central building. While detector output data are only acquired in the central station, 
 %environmental monitoring data are recorded in all three buildings. 

 In addition to the main detector output (strain channel) which contains the majority of the gravitational
 wave information, we need to record a large amount of  monitoring data from the various subsystems. 
 For instance, the fluctuation of laser in the cavity, the electric pick ups in the length sensing 
feedback and error 
 signals, glitches in the auto alignment control system, can all couple and appear in the strain channel.
A number of environmental sensors are installed  to help identify how external disturbances 
couple to the detector output. Magnetic and electric fields can affect the actuators used for
 controlling the mirrors' positions. Wind speed and wind direction sensors gather 
information about the weather condition which might enhance the level of vibrations 
acting on the interferometer. Acoustic signals which could couple to tanks and mirrors 
are measured by microphones. Changes in the temperature, surrounding air pressure and 
vacuum pressure inside the tanks can lead to slow drifts that will show up in the 
strain channel. 
\subsection{Detector characterization.}
The various possible couplings between subsystems and the strain channel require detailed investigation. Detector 
characterization, which is a collaborative work of scientists from different institutions, serves a dual purpose: 

(i) to help experimentalists identify noise sources and improve the performance of the detector,  

(ii) to  provide data quality information and to identify  veto channels\footnote{Channels which proved to be coupled to $h(t)$ and could be used to veto spurious events in 
the strain channel.}.

The data coming from the strain channel and various subsystems is digitized and recorded at sample rates of up to 16384 Hz \cite{DAQ}.  
Then this data is continuously transmitted via a dedicated radio link (32Mbits/sec) to Hanover where the 
data is stored and where analysis of this data can take place.
The data collected at the interferometer is large in volume and complex in nature. 
Therefore, the \geo Online Detector Characterization System (\godcss), whose description is the main 
content of this paper, was developed to efficiently analyze this data online. 
We assume that each data channel
has a noise component which is stationary, but which is contaminated by several non stationarities 
such as transients, fluctuating noise power in given frequency bands, drifting of narrowband 
noise sources, noise coupling between data 
channels, {\it etc}. It is these phenomena that we are interested in. 

\begin{figure}[tbh]
\centerline{\epsfysize=3cm \epsfbox{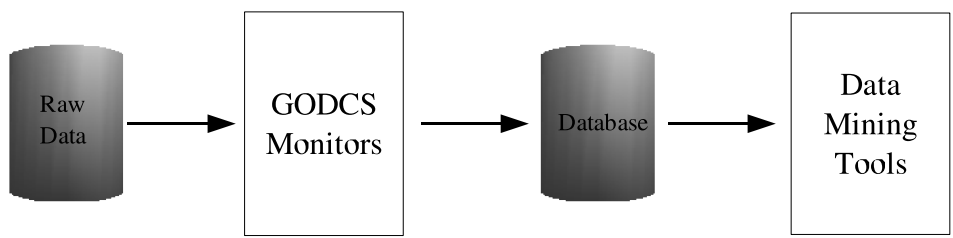}}
\vspace{0.2cm}
\caption{A schematic representation of the data analysis chain in
\godcss. Raw data is processed by \godcs monitors, which produce summary records that are stored in a database. 
The records from the database are further analyzed using data mining tools.}
\label{GODCSfig}
\end{figure}

A schematic representation of \godcs is presented in Figure \ref{GODCSfig}.
The main functional units of \godcs 
are monitors, each of them being an implementation of a well defined data analysis algorithm.    
Since this task is computationally intensive, the system has been designed so that the 
various monitors can be distributed over a cluster of networked computer nodes. 
\godcs has a highly modular architecture which enables monitors to cooperate with each other
by sharing data and data products. Frequently, monitors work in well defined sequences called 
pipelines to carry out a given task. 
Each monitor records the results (if any) of its analysis to a database which is an 
organized store of information. 

The database therefore contains a record of non 
stationarities in a large number of data channels.  
Although data stored in the database is a small fraction of the  
raw data we still require additional tools to display and comprehend
the recorded information. Moreover, since all data channels contain noise, it is 
 certain that some of the results stored in the database are purely false alarms.
Exploration of data recorded to database and extracting the 
useful information is termed as data mining, which is also a part of \godcss.

\subsection{Science runs.}
\geo took data during two short science runs between 2002 and 2004. The main
aims of these runs were to perform astrophysical analysis and to calibrate performance of the detector both in terms 
of sensitivity and stability. \godcs was running online during both data taking runs and 
has shown a very stable and reliable performance. 
The output of \godcs feeds into and impacts gravitational wave data analysis. It serves as a measure of sensitivity of 
the detector and the reliability of the data at any given time and also helps in adding confidence to the results of 
gravitational data analysis. 
For example, the \godcs monitor called {\tt PQMon} \cite{PQMon} was used for vetoing spurious events 
during the search for unmodeled gravitational wave transient signals.
This paper, we believe, will be useful to researchers in the area of gravitational wave 
astronomy as a record of the various analysis and checks carried out to ensure the quality and reliability of the data.

 This paper is structured as follows. 
 In Section~\ref{GEOSoft} we describe the implementation and the overall architecture of \godcss. 
 There we introduce \geopp -- the software on which \godcs is based. Then we outline the 
signal processing tools and main units used in \godcss.  A description of the main \godcs
monitors as well as their implementation is given in the Section~\ref{Monitors}. Finally, 
the data mining process is described in Section~\ref{Triana}. 
We summarize our disscussion and indicate future directions in the concluding section~\ref{Conclusion}.

% DC work ...

%% file: dsp.tex
\subsection{Digital signal processing library in GEO++}
\label{dsp}
The \geopp digital signal processing library is a general purpose library which contains a set of tools needed for data analysis of discrete time series.
In this subsection we will describe the main signal processing units available within \geopp and briefly mention their implementation.

We use fftw3.0 \cite{fftw} for computing discrete Fourier transformations in various frequency based application, for instance, it is used for power spectral density (\psd) estimation. 
Three different methods for estimation of \psd\ are currently implemented in \geopp. Two of them are 
different implementations of Welch overlapped segments averaging (WOSA) method  
and the third one is based on a parametric method using an autoregressive (AR) model \cite{SpecAnal}. 
The WOSA method is very efficient in reducing variance in \psd\ estimation
and the most suitable for long data segments. The estimation of \psd\ based on 
AR model is efficient on short data segments and in absence of periodic signals in the time series. We use window functions in order to reduce the bias caused by a very large dynamical range. For more details on \psd\ estimation we refer reader to \cite{SpecAnal} and references therein.
One can also estimate the noise floor by using running median on the computed
\psd. Running median method is very efficient in removing lines from the \psd\ in addition to the overall smoothening. 
This method is based on substituting the actual value of PSD in each frequency bin by the median value
computed in a narrow frequency band around it.
In addition to the \psd\ we can construct time-frequency maps, and we currently support  
spectrograms and Wigner-Ville maps \cite{warrenbala, tracksearch}.

Linear filters, both IIR (infinite impulse response) and FIR (finite impulse response), are implemented in the time domain. Coefficients for filtering are either supplied by the user or can be constructed internally
using window functions for FIR filters and Butterworth analog filters with bilinear transform for IIR
filters \cite{DSP}.

A resampling routine is used to change the sampling frequency of a discrete time series by
a factor $m/n$, where $m$ and $n$ are integers. The algorithm consists of three
parts: (i) up-sampling (interpolation) by factor $m$ (ii) applying anti-aliasing 
low-pass filter, and (iii) down-sampling by a factor $n$.  The resampled data series is continuous across 
the boundary of contiguous segments, but shifted with respect to the original one. 
The time shift (which we need to account for) is computed.

Finally, there are some other useful tools which are not directly related to signal processing. 
For example we take an advantage of the widely accepted GNU Scientific Library (GSL) \cite{gsl} which contains a large number of very useful routines. We mainly use the GSL library to compute various probability distributions and for the noise generation.

%\section*{References}

%\begin{thebibliography}{99}

%\bibitem{DSP} J.~G.~Proakis and D.~G.~Manolakis {\it Introduction to digital
%signal processing}, Macmillian Publ., 1988

%\bibitem{SpecAnal} D.~B.~Percival and A.~T.~Walden {\it Spectral analysis for 
%physical applications}, Cambridge Univ. Press, 1993.

%\bibitem{numRec}W.~H.~Press, S.~A.~Teukolsky, W.~T. Vetterling and
%B.~P.~Flannery {\it Numerical recipes in C}

%\bibitem{fftw} Webpage {\tt http://www.fftw.org/}
%\end{thebibliography}

%\end{document}

%% file: geopp.tex
\subsection{Frame and MySQL Interfaces}
\label{mysqlframe}
In Section \ref{Intro} we have described the complex nature of the data produced by the \geo instrument.
In addition to storing the raw time-series corresponding to several data channels, we need to store the names of channels, their sampling rates, the coefficients of the various whitening filters if any were applied, {\em etc}.
A standard data format has been agreed to by the various gravitational wave observatories around the world to facilitate easy data exchange. This format is called the frame format \cite{frameformat,framelib}. The frame format is a structured file format which is well suited to store both time-series data as well as meta data that describes how the data was produced. A library of routines to access the data in frames has been developed at VIRGO \cite{framelib}.
%In \geopp the {\tt FrameInputStream} class provides the functionality needed to access the data in frames.

As described in Section \ref{Intro}, the GEO++ monitors send their output results to a database. A database is a convenient
store of complex information and can be thought of as a collection of tables. Each table has a definite structure and in our case represents a particular type of an event such as a glitch or a sudden fluctuation of power in a channel. A table is described by specifying its columns or fields. Each record in the table is represented by a row and will have values corresponding to each column.
As an example, a glitch in any channel of data is recorded to the {\tt glitch} table and can be characterized by the channel in which it occurs, the time of occurrence, the duration of the glitch and its strength. The {\tt glitch} table in the database has therefore at least these columns. Each glitch event is then represented as a single row in this table. It is also useful to link tables together. Very often one needs to know the parameters of an algorithm that detected a glitch event. Therefore in our example the glitch table will have at least one more column which contains a link to other tables in the database which in turn contain information about the parameters used.
The software that manages and manipulates the database is called a database management system (\dbms). There are several \dbms packages available today. The \mysql \dbms \cite{mysql} is one such and is known for its speed and efficiency.
Moreover the software is open source and therefore freely available. A description of the tables and the relationships amongst them will be given in Section \ref{dbasestruct}.

Within \geopp we have software classes that interface with the \mysql and the frame libraries and all access to the raw data and database is routed through these classes.

\subsection{GEO++ Monitors}
\label{geoppmon}
Each detector characterization algorithm is implemented as a \geopp {\em monitor}. 
It is in the development of monitors that the object-oriented features of the \cpp programming language 
\cite{Stroustrup} are deployed. Every monitor is derived from a base abstract class called {\tt BaseMonitor} and 
inherits the functionality of its base class which enables each monitor to access data and store the results 
in a database.  The monitors themselves can therefore focus entirely on implementing the 
desired data analysis algorithm. 
Each monitor requests for  data corresponding to a set of channels and the data is made 
available to it in segments. The segment size and the overlap between segments are determined by the monitor. 
The monitors which are currently in use for the purpose of detector characterization are described in 
Section \ref{Monitors}.

\subsection{GODCS Software Architecture}
\label{geopparch}
As described earlier, each \geopp monitor implements  a well defined algorithm and outputs results to a database. Typically each monitor will be operating on several tens of channels. Moreover, as described in the previous Section there are several monitors which are required to run for the purposes of detector characterization. The resulting computational load and memory requirements are substantial and cannot be carried out on a single workstation.  Therefore the analysis has to be carried out on a cluster of networked workstations.  Such a computer network is known as Beowulf cluster \cite{beowulfWebsite}.

Several parallel programming environments exist which enable the user to distribute the computational load across a Beowulf cluster. One such is the Message Passing Interface or MPI software \cite{MPICH}. In this paradigm processes running on a Beowulf network  cooperate by sending messages to each other and the paradigm is extremely well suited for detector characterization where computation to communication ratio is very large. In essence, a master process distributes data to client processes running on other compute nodes. Each client processes the data and sends the results back to the master which then pushes the output to a  database.

\begin{figure}[tbh]
\centerline{\epsfysize=7cm \epsfbox{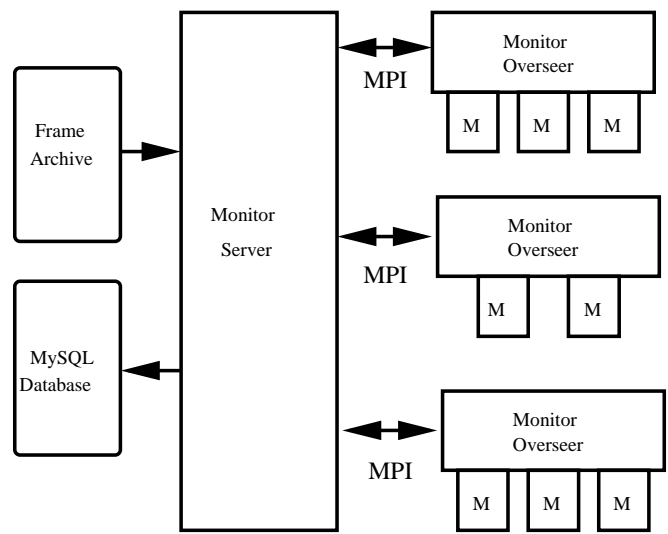}}
\vspace{0.2cm}
\caption{The GODCS analysis pipeline. Each client process consists of a Monitor Overseer and one or more monitors attached to it. The Monitor Overseer handles  all the I/O requests of the monitors. }
\label{godcsFigure}
\end{figure}

We illustrate the GODCS software architecture in Figure \ref{godcsFigure}.
Each client process can be visualised as a Monitor Overseer which controls the execution of the individual monitors.
The Monitor Overseer has the following functionality:
\begin{enumerate}
\item get data from the Monitor Server process,
\item initialize the monitors at the outset and cleanly terminate them  when the analysis has finished,
\item loop over each segment of data and pass control to the monitors in a definite sequence,
\item ensure that the data to be analysed by the monitor is valid and
\item collect the output and pass them to the Monitor Server.
\end{enumerate}
The Monitor Server process has interfaces to the frame data repositories and the MySQL database and provides all the I/O services that the Monitor Overseer requests.

All monitors in a client process see the same copy of the data. Thus normally a monitor would copy the data into its buffers before further processing it. However, in some situations it is actually beneficial for monitors to alter the data available to the  monitors further down in the sequence, {\em e.g.} in many cases the data has to be suitably filtered before a particular algorithm can work efficiently. In such cases one would simply attach the required filter monitor to the sequence at the appropriate place in the monitor sequence. Each sequence of monitors controlled by a single Monitor Overseer process is called a pipeline.

\subsection{Database Structure}
\label{dbasestruct}
Monitors record their output events to a database. Typically each monitor's output is written to a table designed 
specifically for that monitor. These tables in general will be called the {\em event tables}. 
The structure of the table depends on what information the monitor wishes to record about the event.
Since a large number of monitors will be running on several tens of channels it is also extremely useful to keep a record of the various monitors that were run, their parameters and the channels on which they operated upon. Moreover this information must be linked to the actual events stored in the event tables. A relational database readily allows us 
to implement this requirement by allowing fields across tables to be related to each other.

\begin{figure}[tbh]
\centerline{\epsfysize=10cm \epsfbox{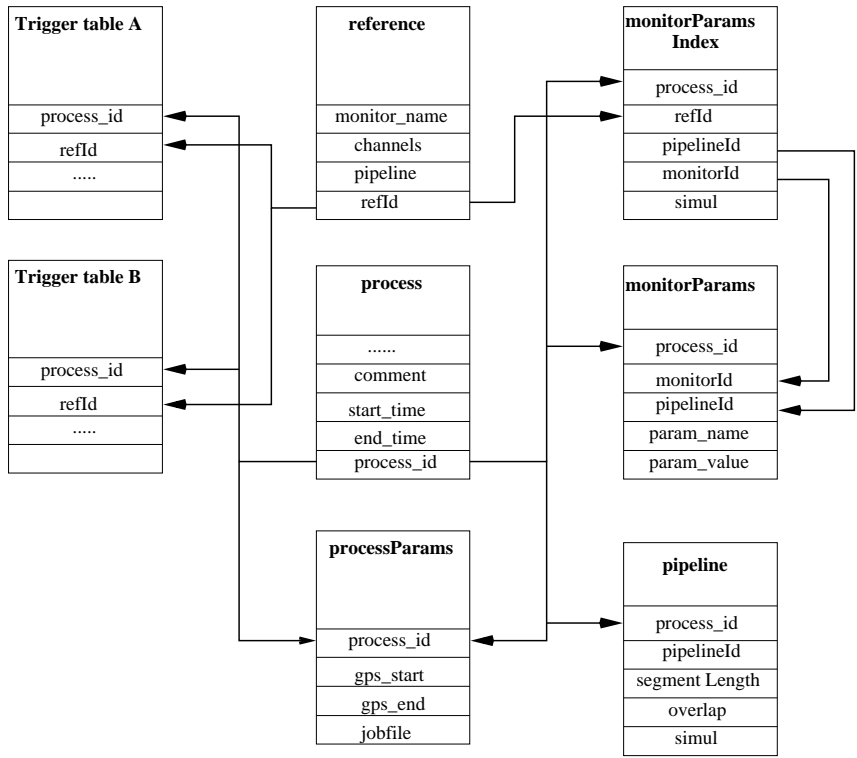}}
\vspace{0.2cm}
\caption{GODCS database structure.}
\label{databaseFigure}
\end{figure}

A  schematic illustration of the database structure used in \geopp is shown in Figure \ref{databaseFigure}. The tables marked as {\tt trigger table A} and {\tt trigger table B} denote event tables. The fields
{\tt process$\_$id} and {\tt refId} fields are common to all event tables. The other tables  store
house keeping information such as the parameters of each monitor when run on a channel.
 The {\tt process} table stores information about the program itself such as the name of the program, its version, start and stop times {\em etc.} There is also an integer field called {\tt process$\_$id} which is generated automatically and is unique in the sense that no two records in the process table can have the same value for
{\tt process$\_$id}. 
For every run a single record is written to the process table. The auto generated value of process\_id is 
then stored along with every record inserted into the database while the program is running.

The {\tt reference} table stores information about the various monitors and the channels they operated upon. 
The primary purpose of this table is to save space in the event tables. Since the names of channels and monitors 
occupy a lot of storage space, each
monitor name, channel combination is mapped to a unique value of {\tt refId} which is a field in the {\tt reference} table. Each monitor then records the appropriate value of {\tt refId} with every 
record in the trigger tables. 
Finally, the various parameters of the monitors are saved in the tables 
{\tt processParams, monitorParamsIndex, monitorParams} and {\tt pipeline}.

%% file: PsdMon.tex
\subsection{PsdMon}\label{PsdMon}
Many of the monitors  described subsequently  require the power spectral density to be computed. Therefore  a monitor which computes an estimate of the power  spectral density is needed and is a cornerstone for most of the analysis. This monitor is called the {\tt PsdMon} and it is usually the first monitor used in pipelines (see Section~\ref{geopparch}). There are several facilities within the Digital Signal Processing core of \geopp for spectral estimation such as Daniell periodogram, Welch periodogram and/or autoregressive method (Section~\ref{dsp}). 

The {\tt PsdMon} monitor computes the one-sided Power Spectral Density using the Welch periodogram 
implemented within the \geopp {\tt PsdStatic} class.  As required by the Welch periodogram, it chops up the 
input data into overlapping segments with a standard 50\% overlap and applies a data window (Hanning) to each
data segment before computing the Fourier transform.  

This monitor has also a facility to estimate the noise floor of the spectrum. 
The term noise floor is defined as a measure  of the underlying level of noise present in the spectrum 
when the various spectral-lines are removed. Its estimation can be done by using a median estimator with 
an appropriate window. 

%% file: LineMonitor.tex
\subsection{LineMon}\label{LineMon}
\begin{figure}
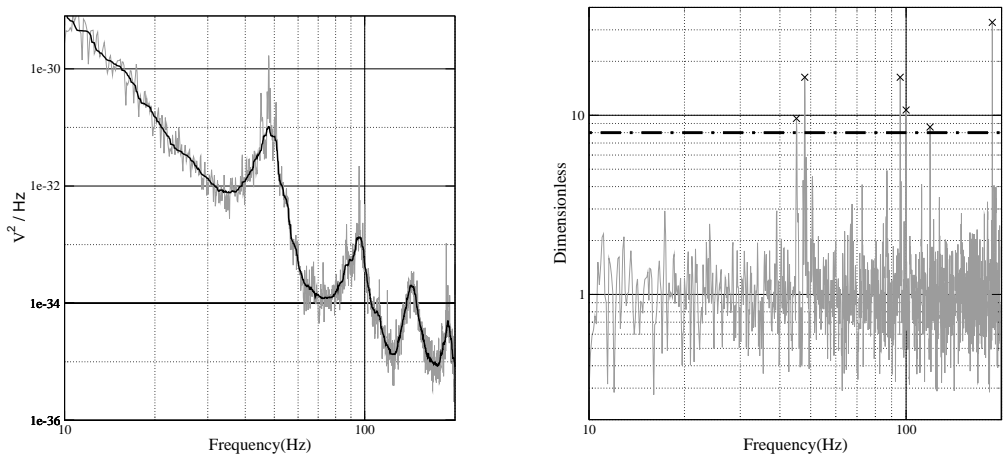

   \centering   
   \includegraphics[width=6cm]{spectrum.eps}
	\hspace{1cm}
   \includegraphics[width=6cm]{normpsd.eps}
   \caption{Example of a spectrum and its noise floor estimator on the left panel (respectively gray and black curve). On the right panel is the normalised spectrum (Equation \ref{eqline}). The normalised spectrum has a known distribution namely a Gamma distribution; a threshold is therefore fixed by a false alarm rate. Each events which crosses the threshold is marked with a cross.\label{fig:linemon:npsd}}
\end{figure}

\begin{figure}
   \centering
   \includegraphics[width=8cm,height=6cm]{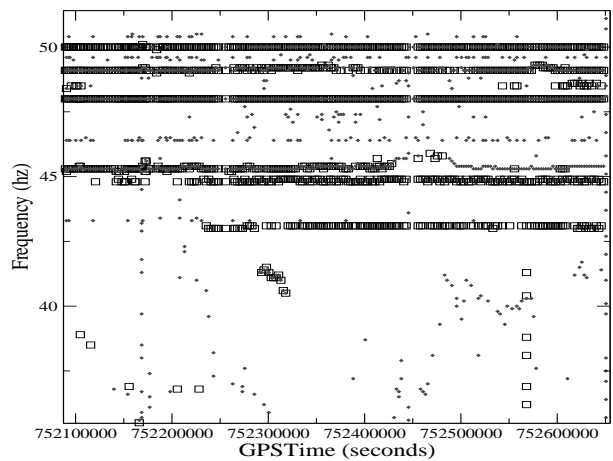}
   \caption{LineMon is applied to successive spectra and the characteristics of the detected lines are stored in the database. 
Therefore various data mining can be done off line. In particular,  a time frequency representation can be extracted as illustrated above: we have superposed the detected events in a narrow frequency band of 15Hz and a time duration of one day; squares and dots correspond to two different channels.   \label{fig:linemon:tf}}
\end{figure}

Each channel has usually a complex spectrum composed of  numerous types of narrow band features (denoted lines in the following).  Lines such as the 50Hz power line and its harmonics, calibration lines injected by an experimenter to assess the sensitivity \cite{Calibration}, violin modes of the test mass suspensions \cite{stefan} and lines from unknown origins,  are indeed present in most of the channels. The motivation for a line detection is to automatically detect the lines especially in the main output channel. The output of the line monitor can be used either by another monitor or by data mining tools to find their origins.  Eventually, the main objective is to remove the lines by modifying the hardware or using signal processing techniques.

Starting from the PSD estimate, we use a detection algorithm which computes a {\it normalised} spectrum $\hat{P}_k$ as follows :
\begin{equation}\label{eqline}
  \hat{P}_k(f) = \frac{P_k(f)}{P^{NF}_k(f)} , 
\end{equation}
where $P_k$ and $P^{NF}_k$ are, respectively, the spectrum and the noise floor estimator as illustrated in left panel on Figure~\ref{fig:linemon:npsd}.
 
In the case of a  white Gaussian noise, the statistics of $\hat{P}_k$ can be derived precisely. It follows a Gamma distribution whose parameters depend only on the number of segments used in the PSD estimation. The probability density function associated with a false alarm rate which is normally a user input parameter gives us the threshold to apply on  $\hat{P}$. Any spectral feature above that threshold is considered to be a spectral line (see Figure~\ref{fig:linemon:npsd} for an illustration). Since lines  can be characterized by a frequency, an amplitude and a frequency width, these quantities  are recorded in a database.

Using information stored in a line-data table,  a time-frequency representation of the data can be derived for the purpose of  data mining, and allows us, in particular, to study the behaviour of drifting lines as illustrated on Figure~\ref{fig:linemon:tf}. A 
detailed description of the {\tt LineMon }  and its application can be found in \cite{linemon-paper}.

%% file: StasMonitors.tex
\subsection{PowerTrackerMon}

The output of a detector is not stationary on the long time scale. This can manifest itself in hourly or daily drifts in level of a noise floor in different frequency bands or in a short transient events. In order to help us in monitoring such changes in the detector's behavior we use {\tt PowerTrackerMon}.

This monitor produces two types of records: the first one 
is a trend information which helps us to study power drift
in several frequency bands; the second type of records is of 
burst type. We record any sudden jump in power, that helps us to identify the source of burst by comparing several 
channels.

{\tt PowerTrackerMon} uses estimation of the noise floor produced by {\tt PsdMon} (see Section~\ref{PsdMon}). First, we compute the power in several frequency bands. The first two frequency bands are 0-1 Hz and 1-10 Hz. Above 10 Hz, frequency bands are defined according to a geometrical progression which factor is defined by the user. 
This monitor records power in each band once in a while (trend information) and if the power in a certain band for a 
given segment has changed in $A$ times. The threshold $A$ can be adjusted for different frequency bands according to $A_k =  \textrm{max} \{ A q^{N/2-k},~1.1\}$, where $k$ is the number of frequency bands (starting at the  
low frequency end) and $N$ is the total number of bands.
Therefore the threshold can be made higher at low frequencies where the time series is noisier and
lower at the high frequency end where the data is usually relatively quiet.

\subsection{SaturationMon}

  {\tt SaturationMon} checks whether the analog-to-digital converter is getting saturated. 
 Essentially it checks whether the absolute value of voltage of the incoming signal is below the maximum
 allowed value. The saturation usually occurs in contiguous data bins. In order to 
reduce the number of records stored in the database, we cluster contiguous triggers into a single event. Each event is characterized by the duration of saturation and the maximum value during that period.

\subsection{InspiralSenseMon}

{\tt InspiralSenseMon} produces one of the figures of merit. 
{\tt InspiralSenseMon} estimates the sensitivity of the detector to an optimally oriented binary system that would produce 
a gravitational wave signal with  signal-to-noise ratio of 8 ($\rho$ in Eq.(\ref{deff})) at the output of the 
matched filter \cite{matchfilt}. 
This monitor helps to assess quite efficiently the performance of the detector.
%: it shows drifts in sensitivity and reflects
%quite well major abnormalities in the detector's behavior.
The measured sensitivity is represented by the effective distance (in kpc) up to which we can observe such an 
inspiralling system and it is computed according to

\be
d = \frac{M^{-1/6}\sqrt{5m_1m_2}}{\rho \pi^{2/3}\sqrt{6}} \left[ \sum_{k=1}^{N} \frac{(k\Delta f)^{-7/3}\Delta f}
{S_k}  \right]^{1/2} 
\frac{G^{5/6}c^{-3/2}M_{\odot}^{5/6}}{1 \textrm{kpc}},\label{deff}%\frac{1}{1.0292712503 \times 10^{11}},\label{deff}
\en   
where $M=m_1+m_2$ is a total mass given in solar masses, $\Delta f$ is a frequency resolution and $S_k$ is the one-sided noise \psd. 
We compute the effective distance
$d$ for the following three binary systems: (i) neutron star -- neutron star with the mass of each component equal to $1.4 M_{\odot}$
(ii) neutron star -- black hole with masses $1.4M_{\odot}$ and $10M_{\odot}$ correspondingly, and, 
(iii) binary black holes system with the mass of each component equal to $10M_{\odot}$. These binaries 
probe the detector's sensitivity at different frequency ranges.  

%% file: balaMonitors.tex
\subsection{Inter Channel Couplings}

In order to improve the sensitivity of the detector it is important to understand how the various noise sources couple into the main gravitational wave channel. Linear couplings are  the simplest and most well-understood. A common measure of linear coupling is the coherence statistic \cite{Chen} and this statistic is implemented as a monitor named {\tt LinearCoherenceMon}. The coherence function or statistic, $c(f)$, computed for the two channels $A(t)$ and $B(t)$, is given as,
\begin{equation}
c(f) = \left|\frac{\sum_{i=1}^N \tilde A_i (f) \tilde B_i^*(f)}{\sqrt{\sum_{i=1}^N|\tilde A_i(f)|^2 \sum_{i=1}^N|\tilde B_i(f)|^2 }}\right|,
\end{equation}
where $\tilde A_i(f)$ and $\tilde B_i(f)$  are Fourier transforms of the $i^{th}$ data segment $A_i(t)$ and $B_i(t)$, respectively, and the summation is over $N$ segments of data and where $f$ is  the frequency of interest.

However, it is likely that several of the noise couplings that are present in the \geo instrument might be non-linear. Quantifying non-linear couplings is  difficult. A commonly used statistic to quantify quadratic non-linear coupling is the bicoherence statistic which is essentially the normalised bispectrum \cite{Swami} . This statistic can be used to measure non-linear couplings between frequencies in the same channel of data or between two different channels in which case it is termed as cross-bicoherence.
The cross-bicoherence function $b(f_1,f_2)$ is defined as,
\begin{equation}
b(f_1,f_2) = \left|\frac{\sum_{i=1}^N \tilde A_i (f_1) \tilde B_i(f_2) \tilde B_i^*(f_1+f_2)}{\sqrt{\sum_{i=1}^N|\tilde A_i(f_1)|^2 \sum_{i=1}^N|\tilde B_i(f_2)|^2 \sum_{i=1}^N|\tilde B_i(f_1+f_2)|^2 }}\right|,
\end{equation}
where the symbols have the same meaning as in the previous equation and $f_1$ and $f_2$ are the frequencies of interest. If $A(t)$  and $B(t)$ represent the same channel of data then the cross-bicoherence is essentially the bicoherence statistic. This statistic is implemented by a monitor called {\tt BicoherenceMon}.

Another useful statistic to measure non-linear coupling is to compute the correlation coefficient between the amplitudes of the Fourier transforms at the frequencies of interest, which we shall call the amplitude coupling coefficient.
We define two variables $x_i = |\tilde A_i(f_1)|$ and $y_i = |\tilde B_i(f_2)|$ where $i$ is the segment number.
Considering the $x_i$ and $y_i$ to be instances of the random variables $x$ and $y$ respectively, the amplitude coupling coefficient $a(f_1,f_2)$ is then defined as,
\begin{equation}
a(f_1,f_2) = \frac{\overline{(x-\overline x)(y-\overline y)}}{\sqrt{\overline{(x-\overline x)^2}} \sqrt{\overline{(y-\overline y)^2}}},
\end{equation} where an overbar denotes an average over the segments.  This statistic is implemented by the monitor called
{\tt AmplitudeCouplingMon}.

All the monitors described in this Section work in a similar way. The user can specify the frequency ranges of interest and the step size. The monitor then computes the statistic at every pair of frequencies within the specified range. 
If the value of the statistic is greater than a user defined threshold then the ``event" is written to the database. An event is characterized by  the pair of frequencies (except in the linear case where there is a single frequency) at which the value of the statistic exceeded the threshold and the value of the statistic itself. Each statistic defined above can be characterized very well if the noise is Gaussian and stationary, which means that we can compute the false alarm for any given threshold on the statistic or vice versa. The thresholds are then fixed by the monitor depending on the  desired false alarm probability.

\subsection{Detecting Glitches}

Glitches are short duration spikes in the data. Glitches can occur due to a variety of reasons and faulty electronics is one of the major causes.  A comparison of glitches in the calibrated and uncalibrated $h(t)$ channel also helps us in identifying any problems in the calibration procedure.
Since glitches are short lived, the best algorithms to detect them work directly in the time domain. However, it is frequently necessary to apply a high pass filter on the data before we apply a time domain glitch detector to the data. This is essentially because the low frequency noise can often be orders of magnitude higher than the noise at higher frequencies. The \geopp monitor that implements a glitch detection algorithm is called {\tt GlitchMon}. The algorithm is quite simple: we identify two time-scales which are the burst time-scale $\Delta t_b$ and the average time-scale $\Delta t_a$ where $\Delta t_b \ll \Delta t_a$. The burst time-scale is typically very small and is normally set to a few samples.  At any given time we compute the average amplitude of the time series over the two time-scales. If the average over $\Delta t_b$ is much larger than the average computed over $\Delta t_a$ then we say that we have detected a glitch. The glitch events are recorded to the glitch table in the database with information about the time of the burst, its duration and its amplitude.

\subsection{Time-frequency methods for detecting bursts}
\label{tfsection}
In addition to simple glitches, the various detector subsystems could be contaminated by more complex bursts of noise.
Pure time domain algorithms for detecting short period bursts are often not sensitive to a broad class of bursts which have their energy concentrated in a relatively small band of frequencies.  Time-frequency algorithms, however \cite{warrenbala, tracksearch, siong, Julien, excessPower}, are  sensitive to precisely such bursts. The basic idea
is to construct a Time-frequency map of the time series data and then to look for features in the map that are not consistent with noise. There are several Time-frequency representations which are useful, the primary ones being the spectrogram and the Wigner-Ville distribution.  Also there are several methods to pick out features in the Time-frequency plane. The two methods that have been implemented in \geopp are {\tt HACRMon}, which implements the Hierarchical Algorithm for Clusters and Ridges \cite{siong} and ({\tt TrackSearchMon}) which implements the Signal Track Search Algorithm \cite{tracksearch}.  

\begin{figure}[tbh]
\centerline{\epsfysize=10cm \epsfbox{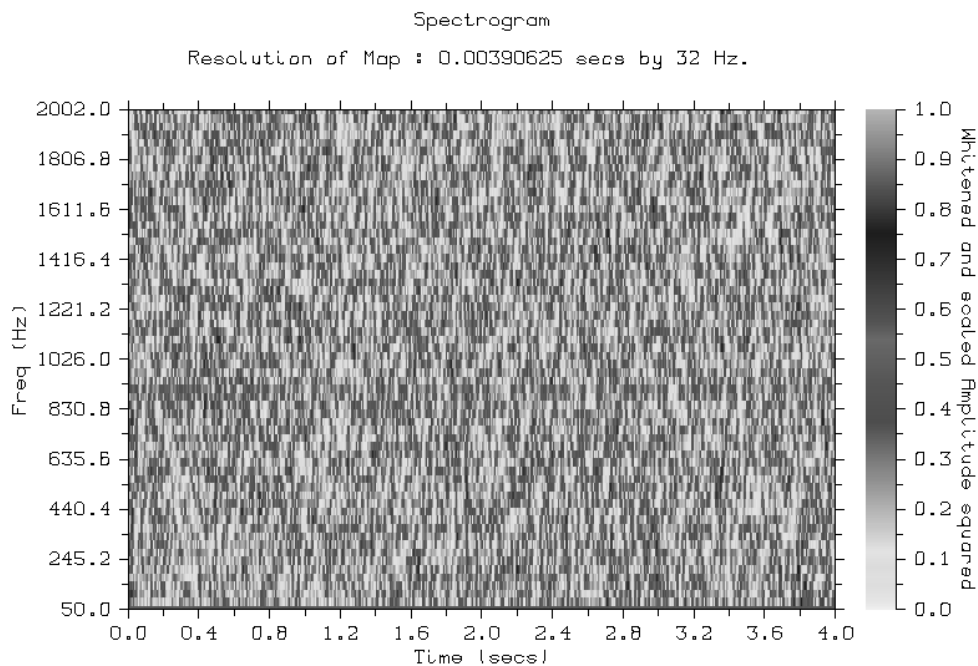}}
\vspace{0.2cm}
\caption{A sample spectrogram of the common mode visibility channel. }
\label{spectrogram}
\end{figure}

A sample spectrogram is illustrated in Figure \ref{spectrogram}. There are several distinct features in the spectrogram shown. The faint horizontal line indicates the presence of a narrowband noise source at around 900 Hz. which could be due to electronic pickup or any resonance in a suspension. However for the purpose of detecting transients we are interested in features which are localised in the Time-frequency plane.  {\tt HACRMon} picks out regions in the Time-frequency plane where an excess power is observed. {\tt TrackSearchMon} is more suitable for narrow curvilinear structures in the Time-frequency plane and utilises not only the amplitude but also the derivatives of the map along the time and frequency axis to infer the presence of signals which leave a curvilinear track on the Time-frequency plane.

%% file: Triana.tex
In earlier Sections we have described how the raw data is processed using several monitors and the results of the analyses are stored as records in tables in the GODCS database. Even though the volume of the database ($\approx$ 1 KB/sec) is a tiny fraction of the volume of the raw data ($\approx$ 1 MB/sec),  it is still a considerable amount of complex information.
Often, we are not just interested in the results from a single table but rather we need to combine the results of several tables. For example if a large number of glitches is observed at certain periods in the main data channel, we might be interested in looking at the interchannel couplings between the main data channel with several other channels at the same periods. This might help us identify the cause of this extra noise.
Also it must be kept in mind that the output of each monitor is bound to contain a few false alarms. This implies that we must further refine the information contained in the database before we can make physically meaningful statements about the state of the detector. This task of extracting useful information from the database is termed as {\em Data mining.}

\begin{figure}[tbh]
\centerline{\epsfysize=10cm \epsfbox{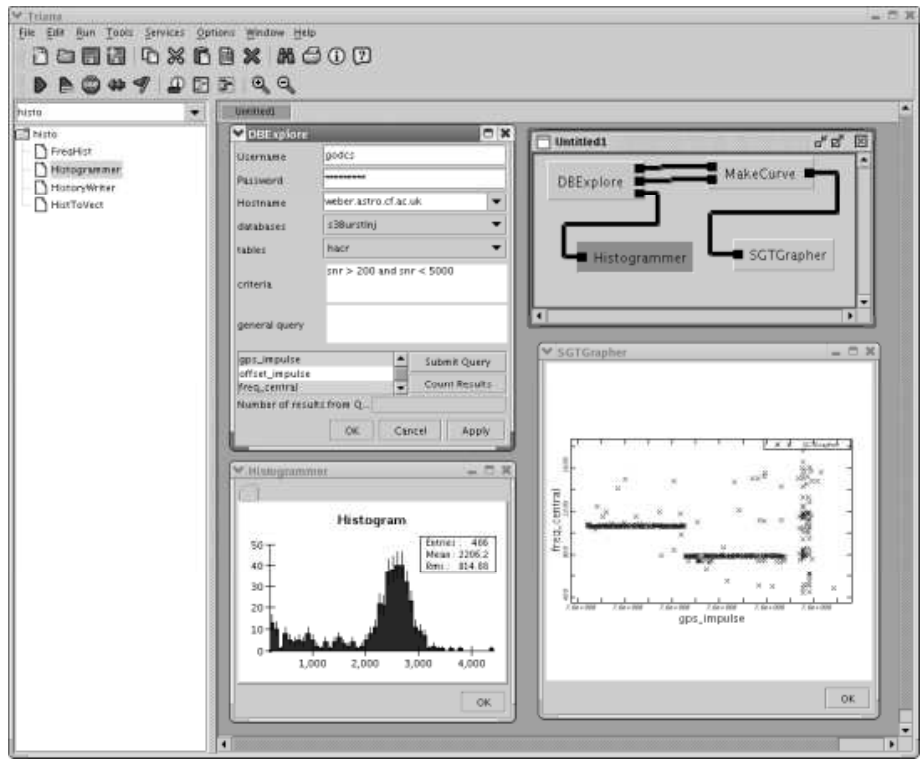}}
\caption{A screenshot of the Triana Software. }
\label{trianaFigure}
\end{figure}

A large part of data mining will be done interactively. For this it is important to have tools to help us graphically visualize the results in a database. We have adopted Triana \cite{TrianaSoft} as our data mining environment. Triana is a Java based visual work flow manager software. It is designed as a collection of pluggable components, or units, which can be assembled together to carry out user defined tasks. A wide variety of Triana units exist and are grouped into toolboxes.
%The signal processing toolbox for example contains components to compute Fourier transforms and carry out time domain filtering. Other components display graphs on the screen and yet other components interface with databases and other file formats. The components can be visually linked to each other via a user friendly GUI.
A screenshot of Triana is shown in Figure~\ref{trianaFigure}. On the top right hand corner of this figure we show a typical data mining workflow constructed out of Triana units. The workflow shown extracts relevant data from the database and converts this numerical information to graphs. In the example shown, a population of burst events extracted from the {\tt HACR} table have been displayed as a pair of graphs, the first of which is a histogram of the signal to noise ratio, while the second is a scatter plot of the events on the time frequency plane.
%The DBExplore component or unit queries a database to get results from the hacr table. Each Triana unit can have user settable parameters and the parameter window for the DBExplore unit is shown on the top left box in the figure. One can choose the name and location of the database as well as the specific table to query. In this example we have chosen to get the time of occurrence, frequency an snr (see section \ref{tfsection}) of events in the hacr table. The snr for all the events  is then passed to the Histogrammer unit which displays a histogram of the snr. The time of occurrence and frequency of the events are sent to the MakeCurve unit which combines these and forwards it to the SGTGrapher unit which displays the events as a time frequency scatter plot. The histogram and the scatter plot are shown in the bottom left and bottom right corners of figure \ref{trianaFigure} respectively.

The discussion above describes only one of the many possible ways in which the database contents can be displayed as graphs. In many cases it is important to look at the results corresponding to different data channels simultaneously. For example it is often very useful to display the burst events for two different channels on the same time frequency scatter plot to check for any relationships between the two event populations. Using Triana's units it is easy to query the database repeatedly for contiguous time intervals and pass the query results to a grapher unit. This enables us to see how event populations change with time and to spot periodic changes in the event rate. Such workflows can be run in the \geo control room to monitor the behaviour of various channels.

Though a direct visual representation of the database results is very useful, it is often necessary to process the database results further. We give a few illustrations below.
\begin{itemize}
\item A coincidence study of burst events in two different channels is useful in determining if the glitch populations in the two channels are related.
\item Tracking glitch rates over a long period of time gives us information about any slow non-stationarities in the data. Often one can notice periodic fluctuations which can be connected to some natural phenomena. For example, it was once observed that the fluctuations of the glitch rate on a time scale of twenty minutes was connected with the air conditioning system which would switch on roughly once every twenty minutes.
\item The output of the {\tt line} table are monitored for extended periods of time to check for drifts in the line frequencies. These could again reveal non-stationarities in the data. Moreover, lines which have a similar drift could be related and, therefore, have a common origin.
\end{itemize}
Automating such tasks is a challenge and  data mining units to carry out these tasks are being developed as generic units in Triana.

%% file: Conclusions.tex
In this paper we have described  the role played by  in detector characterization of the GEO600 instrument. We have summarized the various algorithms that we use to process the various interferometer channels and described how the output of the various monitors  are stored in a database and how these results can be retreived efficiently, displayed  and used to draw conclusions about the state of the interferometer.  The aim of this paper was to give an overview of the various components of GODCS; a detailed description of the various algorithms and monitors will be published elsewhere.

GODCS had been in development for the last three years and we now have a mature software which is being continuously used for detector characterization for both online and offline analysis. The system has proved capable of handling large data rates as well as the large computational load of analysing the various detector channels. The current emphasis is on making the data mining tools more sophisticated so that it is easier to combine the results from various tables in the database.